\documentstyle[epsfig,prl,twocolumn,aps]{revtex} 
\begin{document}
\def\thefootnote{\fnsymbol{footnote}}
\def\com#1{\ \\ \ {\bf \# {#1}}\\ \ }
\def\note#1{$^\surd$\marginpar{\parbox{25mm}
{\raggedright\tiny\scriptsize#1}}}
\def\bort#1{}
\def\nn{\nonumber\\}
\def\eqs#1#2{\mbox{Eqs.~(\ref{#1}) and (\ref{#2})}}
\def\eq#1{\mbox{Eq.~(\ref{#1})}}
\def\be{\begin{equation}}
\def\bea{\begin{eqnarray}}
\def\eea{\end{eqnarray}}
\def\ee{\end{equation}}
\def\inv#1{\frac{1}{#1}}
\def\tr{{\rm tr\,}}
\def\sign{{\rm sign}}
\def\pa{\partial}
\def\goto{\rightarrow}
\def\vek#1{\hbox{\boldmath$#1$}}
\def\slask{\!\!\!\!/}
\def\sslask{\!\!/}
\def\<{\langle}
\def\>{\rangle}
\def\id{\leavevmode\hbox{\small1\kern-3.3pt\normalsize1}}
\def\simleq{\; \raise0.3ex\hbox{$<$\kern-0.75em
      \raise-1.1ex\hbox{$\sim$}}\; }
\def\simgeq{\; \raise0.3ex\hbox{$>$\kern-0.75em
      \raise-1.1ex\hbox{$\sim$}}\; }
\def\g{\gamma}
\def\ve{\varepsilon}
\def\cO{{\cal O}}
\newcommand{\mat}[4]{\left(\begin{array}{cc}{#1}&{#2}\\{#3}&{#4}\end{array}
\right)}
\newcommand{\matr}[9]{\left(\begin{array}{ccc}{#1}&{#2}&{#3}\\{#4}&{#5}&{#6}\\
{#7}&{#8}&{#9}\end{array}\right)}           
\newcommand {\ignore}[1]{}
\newcommand{\nota}[1]{\makebox[0pt]{\,\,\,\,\,/}#1}
\newcommand{\notp}[1]{\makebox[0pt]{\,\,\,\,/}#1}
\newcommand{\braket}[1]{\mbox{$<$}#1\mbox{$>$}}
\newcommand{\Frac}[2]{\frac{\displaystyle #1}{\displaystyle #2}}
\renewcommand{\arraystretch}{1.5}
\newcommand{\noi}{\noindent}
\newcommand{\bc}{\begin{center}}
\newcommand{\ec}{\end{center}}
\newcommand{\epm}{e^+e^-}
\def\ifmath#1{\relax\ifmmode #1\else $#1$\fi}
\def\half{\ifmath{{\textstyle{1 \over 2}}}}
\def\quarter{\ifmath{{\textstyle{1 \over 4}}}}
\def\3quarter{{\textstyle{3 \over 4}}}
\def\third{\ifmath{{\textstyle{1 \over 3}}}}
\def\twothirds{{\textstyle{2 \over 3}}}
\def\fourth{\ifmath{{\textstyle{1\over 4}}}}
\def\sqrthalf{\ifmath{{\textstyle{1\over\sqrt2}}}}
\def\halfsqrthalf{\ifmath{{\textstyle{1\over2\sqrt2}}}}
\def\cl{\centerline}
\def\vs{\vskip}
\def\hs{\hskip}
\def\ra{\rightarrow}
\def\Ra{\Rightarrow}
\def\us{\undertext}
\overfullrule 0pt
\def\lf{\leaders\hbox to 1em{\hss.\hss}\hfill}
\def\21{$SU(2) \otimes U(1) $}
\def\ne{\hbox{$\nu_e$ }}
\def\nm{\hbox{$\nu_\mu$ }}
\def\nt{\hbox{$\nu_\tau$ }}
\def\ns{\hbox{$\nu_{s}$ }}
\def\nx{\hbox{$\nu_x$ }}
\def\Nt{\hbox{$N_\tau$ }}
\def\nr{\hbox{$\nu_R$ }}
\def\O{\hbox{$\cal O$ }}
\def\L{\hbox{$\cal L$ }}
\def\mne{\hbox{$m_{\nu_e}$ }}
\def\mnm{\hbox{$m_{\nu_\mu}$ }}
\def\mnt{\hbox{$m_{\nu_\tau}$ }}
\def\mq{\hbox{$m_{q}$}}
\def\ml{\hbox{$m_{l}$}}
\def\mup{\hbox{$m_{u}$}}
\def\md{\hbox{$m_{d}$}}
\def\neus{\hbox{neutrinos }}
\def\gau{\hbox{gauge }}
\def\neu{\hbox{neutrino }}
\def\eq#1{{eq. (\ref{#1})}}
\def\Eq#1{{Eq. (\ref{#1})}}
\def\Eqs#1#2{{Eqs. (\ref{#1}) and (\ref{#2})}}
\def\Eqs#1#2#3{{Eqs. (\ref{#1}), (\ref{#2}) and (\ref{#3})}}
\def\Eqs#1#2#3#4{{Eqs. (\ref{#1}), (\ref{#2}), (\ref{#3}) and (\ref{#4})}}
\def\eqs#1#2{{eqs. (\ref{#1}) and (\ref{#2})}}
\def\eqs#1#2#3{{eqs. (\ref{#1}), (\ref{#2}) and (\ref{#3})}}
\def\eqs#1#2#3#4{{eqs. (\ref{#1}), (\ref{#2}), (\ref{#3}) and (\ref{#4})}}
\def\fig#1{{Fig. (\ref{#1})}}
\def\partder#1#2{{\partial #1\over\partial #2}}
\def\secder#1#2#3{{\partial^2 #1\over\partial #2 \partial #3}}
\def\bra#1{\left\langle #1\right|}
\def\ket#1{\left| #1\right\rangle}
\def\VEV#1{\left\langle #1\right\rangle}
\let\vev\VEV
\def\gdot#1{\rlap{$#1$}/}
\def\abs#1{\left| #1\right|}
\def\pri#1{#1^\prime}
\def\ltap{\raisebox{-.4ex}{\rlap{$\sim$}} \raisebox{.4ex}{$<$}}
\def\gtap{\raisebox{-.4ex}{\rlap{$\sim$}} \raisebox{.4ex}{$>$}}
\def\lsim{\raise0.3ex\hbox{$\;<$\kern-0.75em\raise-1.1ex\hbox{$\sim\;$}}}
\def\gsim{\raise0.3ex\hbox{$\;>$\kern-0.75em\raise-1.1ex\hbox{$\sim\;$}}}
\def\half{{1\over 2}}
\def\beq{\begin{equation}}
\def\eeq{\end{equation}}
\def\bef{\begin{figure}}
\def\eef{\end{figure}}
\def\bet{\begin{table}}
\def\eet{\end{table}}
\def\bea{\begin{eqnarray}}
\def\ba{\begin{array}}
\def\ea{\end{array}}
\def\bi{\begin{itemize}}
\def\ei{\end{itemize}}
\def\ben{\begin{enumerate}}
\def\een{\end{enumerate}}
\def\ra{\rightarrow}
\def\ot{\otimes}
\def\eea{\end{eqnarray}}
\def\apj#1#2#3{          {\it Astrophys. J. }{\bf #1} (19#3) #2}
\def\app#1#2#3{          {\it Astropart. Phys. }{\bf #1} (19#3) #2}
\def\asr#1#2#3{          {\it Astrophys. Space Rev. }{\bf #1} (19#3) #2}
\def\ass#1#2#3{          {\it Astrophys. Space Sci. }{\bf #1} (19#3) #2}
\def\aa#1#2#3{          {\it Astron. \& Astrophys.  }{\bf #1} (19#3) #2}
\def\apjl#1#2#3{         {\it Astrophys. J. Lett. }{\bf #1} (19#3) #2}
\def\ap#1#2#3{         {\it Astropart. Phys. }{\bf #1} (19#3) #2}
\def\nat#1#2#3{          {\it Nature }{\bf #1} (19#3) #2}
\def\nps#1#2#3{        {\it Nucl. Phys. B (Proc. Suppl.) }{\bf #1} (19#3) #2} 
\def\np#1#2#3{           {\it Nucl. Phys. }{\bf #1} (19#3) #2}
\def\pl#1#2#3{           {\it Phys. Lett. }{\bf #1} (19#3) #2}
\def\pr#1#2#3{           {\it Phys. Rev. }{\bf #1} (19#3) #2}
\def\prep#1#2#3{         {\it Phys. Rep. }{\bf #1} (19#3) #2}
\def\prl#1#2#3{          {\it Phys. Rev. Lett. }{\bf #1} (19#3) #2}
\def\pw#1#2#3{          {\it Particle World }{\bf #1} (19#3) #2}
\def\ptp#1#2#3{          {\it Prog. Theor. Phys. }{\bf #1} (19#3) #2}
\def\Sci#1#2#3{          {\it Science }{\bf #1} (19#3) #2}
\def\jetp#1#2#3{         {\it JETP }{\bf #1} (19#3) #2}
\def\mpl#1#2#3{          {\it Mod. Phys. Lett. }{\bf #1} (19#3) #2}
\def\ufn#1#2#3{          {\it Usp. Fiz. Naut. }{\bf #1} (19#3) #2}
\def\sp#1#2#3{           {\it Sov. Phys.-Usp.}{\bf #1} (19#3) #2}
\def\ppnp#1#2#3{           {\it Prog. Part. Nucl. Phys. }{\bf #1} (19#3) #2}
\def\sjnp#1#2#3{         {\it Sov. J. Nucl. Phys. }{\bf #1} (19#3) #2}
\twocolumn[\hsize\textwidth\columnwidth\hsize\csname @twocolumnfalse\endcsname
\rightline{astro-ph/9803002}
\title{Pulsar Velocities without Neutrino Mass}
\author{D. Grasso${}^1$, H. Nunokawa${}^2$ and J. W. F. Valle${}^1$ }
\address{${}^1$Instituto de F\'{\i}sica Corpuscular - IFIC/CSIC,
             Departamento de F\'{\i}sica Te\`orica \\
             Universitat de Val\`encia, 46100 Burjassot, 
             Val\`encia, Spain\\
         ${}^2$Instituto de F\'{\i}sica Gleb Wataghin, 
          Universidade Estadual de Campinas\\
          13083-970 Campinas, S\~ao Paulo, Brazil}
\date{\today} 
\maketitle
\begin{abstract} 
We show that pulsar velocities may arise from anisotropic neutrino
emission induced by resonant conversions of massless neutrinos in the
presence of a strong magnetic field. The main ingredient is a small
violation of weak universality and neither neutrino masses nor
magnetic moments are required.
\end{abstract}
\vskip2pc]

One of the most challenging problems in modern astrophysics is to find
a consistent explanation for the high velocity of pulsars.
Observations \cite{veloc} show that these velocities range from zero
up to 900 km/s with a mean value of $450 \pm 50$ km/s .  An attractive
possibility is that pulsar motion arises from an asymmetric neutrino
emission during the supernova (SN) explosion. In fact, neutrinos carry
more than $99 \%$ of the new-born proto-neutron star's gravitational
binding energy so that even a $1 \%$ asymmetry in the neutrino
emission could generate the observed pulsar velocities.  To find the
origin of such asymmetry is, however, not a minor task.  One
intriguing possible explanation to this puzzle may reside in the
interplay between the parity non-conservation present in weak
interactions and the strong magnetic fields which are expected during
a SN explosion.  Although several possible realizations of this idea
in the framework of the Standard Model (SM) of particle physics have
been already explored \cite{Chugai,others} a conclusive solution of
the problem is still lacking, and there is some motivation for
looking also at solutions that involve physics beyond the SM.

Recently, several neutrino conversion mechanisms in matter have been
invoked as a possible engine for powering pulsar motion.  Kusenko and
Segr\'e proposed a mechanism \cite{KusSeg96} based on MSW conversions
\cite{MSW}. The idea is based on the observation that the strong
magnetic field present during a SN explosion gives rise to some
angular dependence of the matter induced neutrino potentials
\cite{potentials}.  As a consequence, in the presence of non-vanishing
$\nu_\tau$ mass and mixing the resonance sphere for the
$\nu_e-\nu_\tau$ conversions is distorted. If the resonance surface
lies between the $\nu_\tau$ and $\nu_e$ neutrino spheres, such a
distortion would induce a temperature anisotropy in the flux of the
escaping tau-neutrinos produced by the conversions, hence a recoil
kick of the proto-neutron star.  In order to account for the observed
pulsar velocities the required strength of the dipolar component of
the magnetic field between the two neutrino spheres must exceed
$10^{15}$ Gauss \cite{KusSegC} or even larger \cite{Qian}.  Another
crucial ingredient in this mechanism is the neutrino squared mass
difference, $\Delta m^2 \gsim 10^4 {\rm eV}^2$, which leads to \mnt
$\gsim$ 100 eV or so, assuming a negligible $\nu_e$ mass. This is
necessary in order for the resonance surface to be located between the
two neutrino-spheres.  It should be noted, however, that such
requirement is at odds with cosmological bounds on neutrinos masses
unless the $\tau$-neutrino is unstable.
   
Akhmedov, Lanza and Sciama \cite{ALS} proposed a similar pulsar
acceleration mechanism based on resonant neutrino spin-flavour
precession (RSFP) \cite{RFSP}.  The magnetic field not only affects
the medium properties, as in the matter density in the MSW case, but
also induces the spin-flavour precession through its coupling to the
neutrino transition magnetic moment \cite{TM}. The lowest magnetic
field strength required is $B > 2 \times10^{16}$ Gauss, as long as the
neutrino magnetic moment exceeds $\mu_\nu \simgeq 10^{-15}~\mu_B$.
 
In this letter we investigate the relevance, for pulsar motion, of a
different kind of neutrino conversions not requiring any neutrino mass
nor magnetic moment. The basic mechanism was proposed over ten years
ago in ref. \cite{valle}.  In contrast to the physics of the solar
neutrino problem, the new mechanism was shown to be potentially
relevant for supernova physics. This has been recently studied in more
detail in ref. \cite{NQRV} where stringent limits have been derived.

The simplest underlying particle physics model that realizes this new
conversion mechanism postulates the existence of two new \21 singlet
leptons for each generation of leptons, in such a way that lepton
number symmetry is exact in the Lagrangian \cite{MV}. These extra
states can arise in various extensions of the SM, such as superstring
models \cite{W}. However, the model is very interesting on its own
right, both conceptually as well as phenomenologically \cite{beyond}.
As a result of the postulated lepton number symmetry neutrinos remain
massless to all orders of perturbation even after the gauge symmetry
breaking. However, unlike the situation in the SM 
there is a non-trivial Kobayashi-Maskawa-like mixing in the weak
leptonic charged current \cite{2227}.  The simplest such scheme
contains three two-component gauge singlet neutral leptons $S$ added
to the three right-handed neutrino components $\nu^c$ present in
SO(10). For definiteness we consider this model at the \21 level. The
assumed conservation of lepton number leads to a neutral mass matrix
with the following texture in the basis ($\nu, \nu^c, S$):
\beq
\matr{0}{D}{0}
{D^T}{0}{M}{0}{M^T}{0},
\label{matrix}
\eeq
where the Dirac matrix $D$ describes the coupling between the weak
doublet $\nu$ and the singlet $\nu^c$, and whereas $M$ connects the
singlet states $\nu^c$ and $S$.  It is easy to see that, as expected,
the three conventional neutrinos remain massless, while the other six
neutral 2-component leptons combine into three heavy Dirac fermions.
This model offers a viable alternative to the {\it see-saw} model.

The phenomenological implications of this picture are manifest 
when considering  the resulting charged-current Lagrangian 
in the massless-neutrino sector:
\beq
\label{CC}
{\cal L}_{\rm CC}= \frac{\mbox{i} g}{\sqrt{2}} W_\mu \bar{e}_{a L} 
\gamma_\mu  K_{a i} \nu_{i L} + \mbox{h.c.}~,
\eeq
where $a=e,\mu,\tau$, $i=1,2,3$ and the mixing matrix $K$ is given
as \cite{valle,2227}
\begin{equation}
K = R\,{\cal N},
\end{equation}
For definiteness and simplicity we will confine ourselves to the case
of two neutrinos. In this case $R$ is a $2 \times 2$ rotation matrix,
\begin{equation}
R= \left(\matrix{\cos \theta &\sin \theta \cr 
                - \sin \theta &\cos \theta } \right),
\label{mixing}
\end{equation}
and where the diagonal matrix, 
\begin{equation}
{\cal N} =  \mat{{\cal N}_1} {0} 
                  {0} {{\cal N}_{2,3}},
\label{nonuniversal}
\end{equation}
Note that $K$ is not unitary, since it is a sub-matrix of the full
rectangular matrix including also the heavy states \cite{2227}. The
matrix ${\cal N}$ describes the effective non-orthogonality of the two
neutrino flavours, i.e., $\langle\nu_e|\nu_{\mu,\tau}\rangle \equiv -
\sin\theta \cos\theta ({\cal N}_1^2 - {\cal N}_{2,3}^2)$. The
non-diagonal elements of the matrix $K$ cannot be rotated away through
a redefinition of the massless-neutrino fields. In this way a
non-trivial mixing arises among the massless neutrinos.  

The corresponding form of the neutral-current Lagrangian for the
massless-neutrino sector is \cite{valle,2227}
\beq
\label{NC}
{\cal L}_{\rm NC}= \frac{\mbox{i} g}{2 \cos \theta_{W}} Z_\mu P_{ij} 
\bar{\nu}_{i L} 
\gamma_\mu  \nu_{j L}\:,
\eeq
where 
\beq
P = K^{\dagger} K = {\cal N}^2~.
\eeq
The matrix $P$ is diagonal but generation-dependent, signalling the
violation of weak universality.  It is also convenient to define
\begin{equation}
{\cal N}_i^2 \equiv (1+h_i^2)^{-1}, \,\,\,\,\,\, i=1,2(3),
\label{hdef}
\end{equation}
where the $h_i$ parameters reflect the deviation from the {\sl
standard} neutrino coupling.  Since no oscillations between two
strictly massless neutrinos can develop in vacuum, it follows that
laboratory limits on the leptonic mixing angle $\theta$ are very weak.
However, it will be sufficient for our purposes to assume that the
mixing angle $\theta$ is very small.  In this way we have $\nu_i \sim
\nu_a\ [a=e,\mu (\tau)]$, so that $h^2_i \sim h^2_a$.  The parameters
$h_i^2$ are also constrained experimentally.  There have been
extensive experimental studies of the constraints on $h_a^2$. For the
third generation one can still allow $h_\tau^2$ values in the range of
a few percent \cite{univ}, whereas the constraints on $h_e^2$ and $h^2
_{\mu}$ are more stringent. For this reason we consider from now on the
case of \ne \nt conversions.


In the flavour eigenstate-type basis ${\tilde \nu_a}$ defined in
ref. \cite{valle}, the Schroedinger evolution equation describing the
neutrino propagation in matter can be written as
\begin{equation}
{i{d \over dr}\left(\matrix{
\tilde{A}_e \cr\ \tilde{A}_\tau\cr }\right)=
 \sqrt{2} G_F {\rho \over m_N} \left(\matrix{ \tilde{H}_{e} &
 \tilde{H}_{e\tau} \cr
 \tilde{H}_{e\tau} & \tilde{H}_{\tau} \cr}
\right)
\left(\matrix{
\tilde{A}_e \cr\ \tilde{A}_\tau \cr}\right) },
\label{evolution1}
\end{equation}
where $\tilde{A}_{e,\tau}$ are the neutrino amplitudes, $G_F$ is the
Fermi constant, $\rho$ is the matter density, and $m_N$ is the nucleon
mass. The entries of the evolution Hamiltonian are now given by
\cite{valle,NQRV}
\begin{eqnarray}
\label{hamilt}
\tilde{H}_{e} &  = & Y'_e({\cal N}_e c^2+{\cal N}_\tau s^2)^2-
\frac{1}{2}Y'_n({\cal N}_e^2c^2+{\cal N}_\tau^2s^2), 
\nonumber \\
\tilde{H}_{e\tau} & = & 
[Y'_e({\cal N}_e c^2+{\cal N}_\tau s^2)-\frac{1}{2}Y'_n({\cal N}_e
+{\cal N}_\tau)]({\cal N}_\tau-{\cal N}_e)sc, \\
\tilde{H}_{\tau} & = & Y'_e s^2c^2({\cal N}_\tau^2-{\cal N}_e^2)
-\frac{1}{2}Y'_n({\cal N}_e^2s^2+{\cal N}_\tau^2c^2) \nonumber
\end{eqnarray}
where
\be
Y'_e \equiv Y_e - Y^0_e \cos\phi \qquad 
Y'_n \equiv Y_n + Y^0_e \cos\phi
\ee
where $Y_e\equiv{n_p \over n_p+n_n}$ and the superscript zero refers to
the contribution from the lowest Landau level \cite{foot}.  In
\eq{hamilt} we used the shorthand notation $s \equiv
\sin\theta$ and $c \equiv \cos\theta$. The effect of the magnetic field 
enters through the neutrino potentials \cite{potentials} that affect
the entries of the Hamiltonian. This can be parametrized by defining
effective electron and neutron fractions $Y'_e$ and $Y'_n$ which now
contain an anisotropic term $\cos\phi$ where $\phi$ is the angle
between the neutrino propagation direction and ${\vec B}$.  Using
\eq{hamilt} the resonance condition, ${\tilde V}_e = {\tilde V}_\tau$,
reads
\be
\label{reson1}
Y'_e = \eta Y'_n
\ee
or, in a more transparent form, 
\be
\label{reson2}
Y_e \left(1 + \lambda_e \cos\phi \right) = \eta Y_n \left(1  - \lambda_e 
\frac{Y_e}{Y_n} \cos\phi \right)~.
\ee
where $ \lambda_e \equiv n_e^0/n_e$ where the superscript zero again
the contribution to the electron fraction coming from the lowest
Landau level. The parameter $\eta$ defined as
\be
\eta \equiv \frac 1 2 \left(h_\tau^2 - h_e^2\right)~. 
\ee
measures the deviation from lepton universality which, for the
$(\nu_e,\nu_\tau)$ system may reach $\eta \sim 10^{-2}$ or so
\cite{univ}.

In order to establish that massless neutrino conversions can play a
role explaining the origin of pulsar velocities, we need to verify
that the resonance condition \eq{reson1} can indeed be fulfilled in a
SN  environment between the $e$ and the $\tau$ neutrino-spheres.
The mean resonance position is obtained by averaging \eq{reson2}
over $\phi$, giving
\be
\label{reson0}
Y_e = \eta Y_n~.
\ee
Apart from small ${\bf B}$ induced corrections to $Y_e$, the condition
\eq{reson0} coincides with the free-field resonance condition given in 
ref. \cite{NQRV} which allows us to apply here some of the arguments
used there. As $Y_n \approx 1$ in the core of the SN, we see that
condition in \eq{reson0} can be fulfilled for experimentally allowed
$\eta$ values if $Y_e \lsim 10^{-2}$.  This is indeed possible close
to the neutrino-spheres as a consequence of the strong deleptonization
taking place in that region during the Kelvin-Helmholtz cooling phase.
A rough estimate of the value of $Y_e$ between the neutrino-spheres
can obtained following ref. \cite{NQRV}. From the approximate chemical
equilibrium for $e^-$, $p$, $n$, and $\nu_e$, we have
$\mu_{e^-}+\mu_p\sim \mu_n$, where $\mu_{e^-}$ is the electron
chemical potential and we have set $\mu_{\nu_e}\sim 0$. Using
Boltzmann statistics for non-relativistic nucleons and the above
chemical equilibrium condition , we can write
\begin{equation}
Y_e \sim \frac{1}{\exp\left({\mu_{e^-} / T}\right) + 1},
\end{equation}
neglecting the neutron-proton mass difference.

The chemical potential for relativistic and degenerate electrons near
the neutrinosphere is given by
\begin{equation}
\label{mupot}
\mu_{e^-}\approx(3\pi^2n_e)^{1/3}\approx 51.6(Y_e\rho_{12})^{1/3}\ \mbox{MeV},
\end{equation}
where $\rho_{12}$ is the matter density in units of $10^{12}$ g
cm$^{-3}$.  For typical conditions between the two neutrino-spheres,
$T\sim 4$ MeV and $\rho_{12}\sim 1\div 10$ we find $Y_e\sim 5\times
10^{-2} \div 5\times 10^{-3}$.  These analytical results are in good
agreement with numerical SN models.  Therefore, we can expect resonant
massless-neutrino conversion to occur between the two neutrino-spheres
for a range of values of the parameter $\eta$ which is not
experimentally excluded.

The adiabaticity of resonant massless-neutrino conversions can be easily 
verified by looking at the probability for $\nu_e\leftrightarrow\nu_\tau$
and $\bar\nu_e\leftrightarrow\bar\nu_\tau$ conversions, given by \cite{NQRV}
\begin{eqnarray}
\label{LZ}
  P & = & 1 - 
\exp\Biggl(-\frac{\pi^2}{2}\frac{\delta r}
{L_{m}^0} \Biggr) \\
          & \approx &  1 - \exp\left[
-32 \rho_{12}(r_0) \Biggl( \frac{\eta}{10^{-2}} \Biggr)
      \Biggl(\frac{h_{Y_e}(r_0)}{1 \mbox{cm}} \Biggr)\sin^22\theta 
                 \right]~. \nonumber
\end{eqnarray}
Resonant conversion is adiabatic if $P \simeq 1$, hence whenever 
the resonance width $\delta r =  2 h_{Y_e} \sin 2\theta$ is larger than
the conversion length at resonance $L_{m}^0$. From \eq{LZ} we see that 
such a condition is fulfilled, under the conditions present between the 
neutrino-spheres, if $\sin^2 2\theta \simleq 10^{-6}$. 

The anisotropy in the total momentum of the escaping neutrinos can be
computed as in \cite{KusSeg96,Qian} starting from the resonance
condition \eq{reson2}.  Since, as we discussed $Y_e \ll Y_n$ in the
region of our interest, we neglect the second term on the right side
of \eq{reson2}.  Following \cite{KusSeg96} we parametrize the
resonance surface equation as follows
\be
\label{ressur}
r(\phi) = r_0 + \delta \cos\phi
\ee      
where $r_0$ is the radius of the free-field resonance sphere defined
by \eq{reson0}.  The displacement $\delta$ is found by subtracting the
resonance condition computed for $\phi = \pi$ with the same computed
in $\phi = - \pi$. We find
\be
\label{delta1}
Y_e(r+\delta) - Y_e(r-\delta) \simeq \frac{dY_e}{dr} 2\delta =
2 \lambda_e Y_e(r_0)~. 
\ee
Defining the $Y_e$ variation scale height by
\be
\label{hY}
h_{Y_e} \equiv \left( \frac{d\ln Y_e}{dr}\right)^{-1}
\ee
we can rewrite \eq{delta1} in the more compact form
\be
\label{delta2}
\delta = \lambda_e h_{Y_e}(r_0)~.
\ee
The deformation of the resonance sphere implies an angle dependence in
the temperature of the escaping \nt's, hence to an asymmetry in the
momentum that these neutrinos carry away from the SN.  This is given
by \cite{Qian}
\be
\label{kick1}
\frac{\Delta k}{k} \approx \frac{1}{6} \frac
{\int_0^\pi F_{\nu} \cos\phi \sin\phi d\phi}
{\int_0^\pi F_{\nu} \sin\phi d\phi} \simeq
\frac{2}{9} \delta   h_T^{-1}  
\ee
where the temperature variation scale height $h_T$ is defined in
analogy to the definition of $h_{Y_e}$ given in \eq{hY}.  The factor
$1/6$ in \eq{kick1} accounts for the fact that, out of the six
neutrino and anti-neutrino species, we are assuming that only the
$\nu_\tau$ carries a momentum anisotropy.  We note, however, that an
extra factor of two may be gained in \eq{kick1} if, at the same time,
${\bar\nu}_\tau$ emission suffers resonant conversion between ${\bar
\nu}_e$ and ${\bar \nu}_\tau$ spheres.  This is indeed possible as the
resonance condition \eq{reson1} for $\nu_e-\nu_\tau$ and ${\bar
\nu}_e-{\bar \nu}_\tau$ conversions coincide. This is a characteristic 
feature of this mechanism \cite{valle} which is in sharp contrast with
the case of MSW conversions \cite{MSW}, where either neutrinos or
anti-neutrinos can resonantly convert, but not both.

By substituting \eq{delta2} in \eq{kick1} we finally get
\be
\label{kick2}
\frac{\Delta k}{k} \approx \frac{2}{9}\ \frac{h_{Y_e}}{h_T}\ \lambda_e~.
\ee
Numerical SN simulations \cite{Susuki} typically give ${h_{Y_e}}/{h_T}
\approx 1$ between the two neutrino-spheres.  Hence, we see from
\eq{kick2} that in order for massless neutrino resonant conversions 
to account for the observed pulsar velocities one needs $\lambda_e
\approx 5\times 10^{-2}$ in the region between the two
neutrino-spheres.  Using the expression for the polarization
$\lambda_e$ given in \cite{NSSV} we determine the required value of
the magnetic fields strength as $\approx 10^{15}$ Gauss, which seems
reasonable from the astrophysics point of view \cite{TD}, and lower
than required in the MSW \cite{KusSegC} and RSFP
\cite{ALS} neutrino conversion mechanisms.  

This is illustrated in Fig. 1 for three different choices of $\rho
Y_e$ at resonance consistent with SN models. The three lines
correspond to $\rho Y_e$ values (in g/c.c.) $5\times10^{9}$ (solid),
$1\times10^{10}$ (dashed) and $5\times10^{10}$ (dotted) at resonance.
In Fig. 1 we assume that both $\nu_\tau$ and ${\bar \nu}_\tau$ carry a
momentum anisotropy.  The kick velocity values given in the right
ordinate assume a pulsar mass of 1.4 $M_\odot$ and a total energy
released by all neutrino species of $3 \times 10^{53}$ erg.

\bef
\centerline{\protect\hbox{
\psfig{file=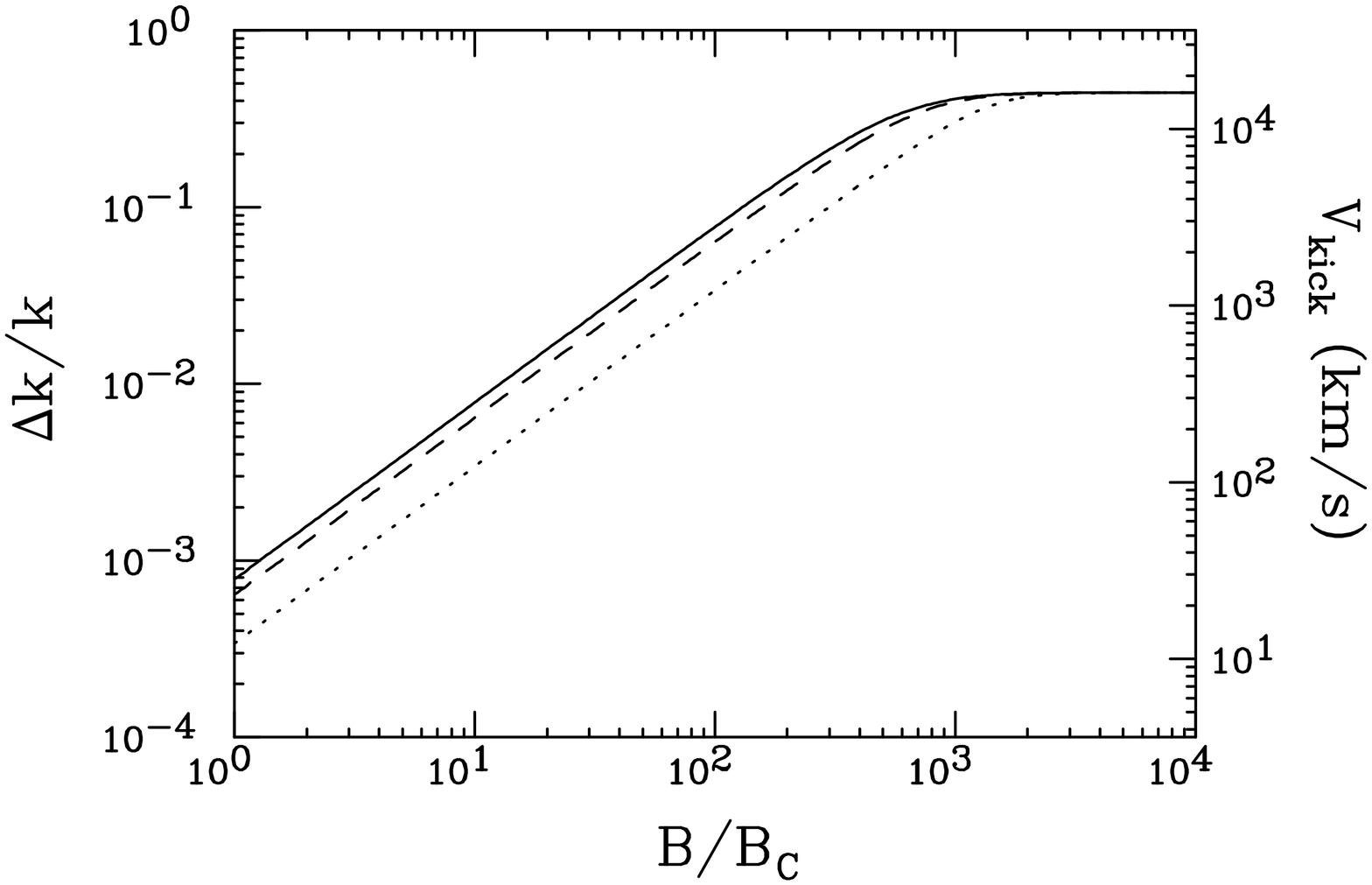,height=6cm,width=7.0cm,angle=90}}}
\caption{Magnitude of kick velocity versus magnetic field, in 
units of the critical field $B_c = 4.4 \times 10^{13}$ Gauss.}
\eef
\noindent

In short we have proposed a viable scheme for generating pulsar
velocities from anisotropic neutrino emission induced by resonant
conversions of massless neutrinos in the presence of a strong magnetic
field. Neither neutrino masses nor magnetic moments are required and
the parameters required are totally consistent with cosmology
(e.g. nucleosynthesis) and astrophysics (e.g. supernova physics).
However, our proposal rests on the idea that there is a small
violation of universality in the weak interaction. Again, we are
safely within the presently allowed region of parameters. However,
this mechanism would become inconsistent should the sensitivity of
weak universality tests improve significantly.

\noindent{\bf Acknowledgements:}
This work was supported by DGICYT grant PB95-1077, by a CICYT-INFN
grant and by the TMR network grant ERBFMRXCT960090 of the European
Union.  H. N. was supported by a FAPESP. We thank A. Rossi for
discussions.


\begin{references}
\bibitem{foot}
In \eq{hamilt} we neglected the neutrino contribution to the
potentials in  between the two neutrino-spheres.
\end{references}
\end{document}